\documentclass[review]{elsarticle}

\usepackage{lineno,hyperref}

\usepackage{amssymb}
\usepackage{bbold}
\usepackage{dsfont}
\usepackage{bbm}
\usepackage{graphicx}
\usepackage{graphicx}
\usepackage{amssymb}
\usepackage{tabularx}
\usepackage[utf8]{inputenc}
\usepackage[english]{babel}
 \usepackage{amsthm}
\usepackage{amsmath,amsfonts}
\usepackage[usenames, dvipsnames]{color}
\usepackage{dsfont}
\usepackage{bbm}
\usepackage{breqn}
\usepackage{amssymb}
\modulolinenumbers[5]

\journal{arXiv} %Journal of \LaTeX\ Templates}

\begin{document}

\begin{frontmatter}

\title{On spatial and spatio-temporal multi-structure point process models%\tnoteref{mytitlenote}
}



\author[mymainaddress]{Morteza Raeisi\corref{mycorrespondingauthor}}
\cortext[mycorrespondingauthor]{Corresponding author}
\ead{morteza.raeisi@univ-avignon.fr}
%\ead[url]{https://math.univ-avignon.fr/annuaire/}
\author[mymainaddress,mysecondaryaddress]{Florent Bonneu}
\author[mysecondaryaddress]{Edith Gabriel}

\address[mymainaddress]{LMA EA2151, Avignon University, F-84000 Avignon, France}
\address[mysecondaryaddress]{INRAE, BioSP, F-84914, Avignon, France}

\begin{abstract}
Spatial and spatio-temporal single-structure   point process models are widely used  in epidemiology, biology, ecology, seismology \dots. However, most natural phenomena present multiple interaction structure or exhibit dependence at multiple scales in space and/or time, leading to define new spatial and spatio-temporal multi-structure point process models. In this paper, we investigate and review such multi-structure point process models mainly based on Gibbs and Cox processes.
\end{abstract}

\begin{keyword}
Spatio-temporal point processes\sep Cox process\sep  Gibbs process \sep Multi-scale process  \sep Multi-structure process.
\MSC[2010] 60G55\sep 62M30 \sep 62H11
\end{keyword}

\end{frontmatter}

%\linenumbers

\section{Introduction}

Fundamental concepts of the theory of point processes emerged from life tables, renewal theory and counting 
problems \cite{daley2003}. The modern theory has mainly been developed between 1940's and 1970's (see e.g. the monographs by Palm~\cite{palm1943}, Feller~\cite{feller1950}, Bartlett~\cite{bartlett1954},  Mat{\'e}rn~\cite{matern1960} and Cox~\cite{cox1955,cox1962})
and is linked to nonlinear techniques in stochastic process theory~\cite{bartlett1955,bosq1998}. From 1980's spatial and spatio-temporal point processes have then become a subject on their own right. Today, they cover a plethora of applications in ecology, forestry, astronomy, epidemiology, seismology, fishery\dots

Spatial (and spatio-temporal) point process data are a collection of points for which locations (and times) of occurrence have been observed in a specified spatial region (and temporal period). Usually, the terms  \textit{points} and \textit{events} are respectively used for arbitrary locations and for observations.
The main goals in the analysis of point patterns concern the specification of intensity variations (first-order moment), interaction between events (second-order moment) and model identification for the underlying process.
Processes are often classified into three classes of interaction structure~\cite{diggle1983}:
\begin{itemize}
  \item \textit{randomness}: In the absence of any interaction between events, a point pattern is said Completely Spatially (or Spatio-Temporally) Random in the sense that the probability that an event occur at any point is equally likely to occur anywhere within a bounded region and that its location (and time) is independent of each any other event. This property provides the standard baseline against which point patterns are often compared. The simplest and most fundamental point process for modelling a complete random distribution of points is the Poisson point process~\cite{kingman1993,kingman2006}. It is used as null hypothesis for statistical test of interaction~\cite{diggle2003,illian2008}.
  \item \textit{clustering} or \textit{aggregation}: In a clustered distribution, events tend to be closer than would be expected under complete randomness. Clustered patterns are mainly modelled by Cox processes~\cite{cox1972}, in particular log-Gaussian Cox processes~\cite{moller1998,brix2001a,brix2001b,diggle2013a}, Poisson Cluster processes~\cite{neyman1958,brix2002,gabriel2014} and Shot-Noise Cox processes~\cite{brix2000,moller2004,moller2010}.
  \item \textit{inhibition} or \textit{regularity}: In a regular distribution, events are more evenly spaced than would be expected under complete randomness. This structure can be modelled by 
      Strauss processes~\cite{strauss1975,cronie2015},     
      Mat{\'e}rn hard core processes~\cite{matern1960,gabriel2013} or determinantal point processes~\cite{macchi1975, lavancier2015}.   
\end{itemize}
Gibbs processes~\cite{ruelle1969,preston1976,dereudre2019} offer a large class of models which allow any of the above interaction structure.

These single-structure point process models are too simplistic to describe phenomena with interactions at different spatial or spatio-temporal scales.
That is for instance the case of seismic data as the different sources of earthquakes (faults, active tectonic plate and volcanoes) produce events with different displacements~\cite{siino2017} and can be seen as the superposition of background earthquakes (which are distributed over a large spatio-temporal scale with low density) and clustered earthquakes
(which are distributed over a small spatio-temporal scale with high density) \cite{pei2012}.
Such multi-structure phenomena motivate statisticians to construct new spatial point process models, e.g. in ecology~\cite{levin1992,wiegand2007,picard2009}, in epidemiology~\cite{iftimi2017} and in seismology~\cite{siino2017,siino2018}, mainly based on Gibbs processes, but not only~\cite{lavancier2016}.
There are very few spatio-temporal models:~\cite{gabriel2017} and \cite{raeisi2019} modeled the multi-scale spatio-temporal structure of forest fires occurrences by log-Gaussian Cox processes (LGCP) and multi-scale Geyer saturation process respectively, \cite{iftimi2018} developed a multi-scale area-interaction model for varicella cases and \cite{illian2012b} modelled the locations of muskoxen herds by LGCP with a constructed covariate measuring local interactions.

In the spatial point processes literature, three general approaches are considered for constructing multi-structure point process models:  hybridization~\cite{baddeley2013}, thinning and superposition~\cite{chiu2013}. Hybridization consists in combining two or more point process models~\cite{baddeley2015}. Spatial hybrids of Gibbs models are defined in~\cite{baddeley2013} and hybrids of area-interaction potentials in~\cite{picard2009}. Extension of the hybridization approach to  the spatio-temporal framework has recently been considered in~\cite{iftimi2018,raeisi2019}. Thinning consists in deleting points of a point process according to some probabilistic rule which is either independent or dependent of thinning other points~\cite{chiu2013}. This operation allows to get point processes with inhibition at small scales and attraction at large scales~\cite{andersen2016,lavancier2016}. Superposition of several processes is the union of the points of each process. It can be useful to model multi-scale clustered processes~\cite{wiegand2007}.

 In this paper, we give a thorough overview of available methods and models for spatial and spatio-temporal  multi-structure point process data.
 In Section~\ref{sec:aboutPP}, we review the required preliminaries which include definitions and properties of point processes and single-structure  models. In Section \ref{sec:methods}, we investigate the spatial and spatio-temporal multi-structure  point process models based on Gibbs and Cox processes and other methods for introducting  new  multi-structure models. Finally, Section~\ref{sec:discussion} provides concluding remarks and discusses directions for future research.

\section{Inhomogeneity and structures in point patterns}
\label{sec:aboutPP}

\subsection{Definitions}

We consider a finite spatial or spatio-temporal point process $X$ observed in ${\cal W}$, where ${\cal W}$ denotes either a spatial region $W \subset \mathds{R}^d$ or a spatio-temporal region $W \times T \subset \mathds{R}^d \times \mathds{R}$.
We denote $\mathbf{x}$ a realization of the point process, i.e. a collection of events $\lbrace x_i \rbrace_{i=1,\dots,n}$ (or $\lbrace (x_i, t_i) \rbrace_{i=1,\dots,n}$) $\subset {\cal W}$. Let $\xi$ be any point in ${\cal W}$.
We refer to~\cite{daley2003,chiu2013} (resp.~\cite{diggle2010,diggle2013b,gonzalez2016}) for more formal definitions of spatial (resp. spatio-temporal) point processes. Without loss of generality, we set $d=2$ throughout this paper.
The main characteristics driving the spatial (resp. spatio-temporal) distribution of points are the \textit{intensity function}, which governs the univariate distribution of the points of $X$, and the \textit{pair correlation function}, which governs the bivariate distribution of the points of $X$, i.e. the interaction between events. In the following we remind some definitions and properties when $X$ is a spatial or a spatio-temporal point process.

Campbell's theorem~\cite{chiu2013} relates the expectation of a function, $h$ assumed to be non-negative and measurable, summed over a point process $X$ to an integral involving the mean measure of the point process~:
$$\mathds{E} \left\lbrack  \sum_{\xi_1,\dots,\xi_k \in X}^{\neq} h(\xi_1,\dots,\xi_k) \right\rbrack
= \int \dots \int h(\xi_1,\dots,\xi_k) \lambda^{(k)}(\xi_1,\dots,\xi_k)
\Pi_{i=1}^k \text{d}\xi_i,$$
where $\xi_i \in {\cal W}$  and $\lambda^{(k)}$, $k \geq 1$, are the product densities. For a simple point process, i.e. $\xi_i \neq \xi_j$ for $i \neq j$, if they exist, the product densities are related to the counting measure $N$ in infinitesimal spatial or spatio-temporal regions d$\xi_1,\dots,$d$\xi_k \subset {\cal W}$, around $\xi_1,\cdots,\xi_k$, with volumes $|d\xi_1|,\cdots,|d\xi_k|$~: $\mathds{P} \left\lbrack N(\text{d}\xi_1)=1, \dots, N(\text{d}\xi_k)=1 \right\rbrack = \lambda^{(k)}(\xi_1,\dots,\xi_k) \Pi_{i=1}^k \text{d}\xi_i$.
Thus, the intensity function is related to the expected number of points in infinitesimal regions $$\lambda(\xi) = \lambda^{(1)}(\xi) = \lim_{|\text{d} \xi| \to 0} \frac{\mathds{E} \lbrack N(\text{d} \xi) \rbrack}{|\text{d} \xi|}$$
and the pair correlation function is defined by
\begin{equation}
    \label{eq:pcf}
    g(\xi_i,\xi_j) = \frac{\lambda^{(2)}(\xi_i,\xi_j)}{\lambda(\xi_i) \lambda(\xi_j)}.
\end{equation}
A point process is \textit{homogeneous} when its intensity is constant, $\lambda(\xi) = \lambda,$ $\forall \xi$, \textit{inhomogeneous} otherwise. In practice, the inhomogeneity is often driven by environmental covariates and we account for them by using parametric models for the intensity function~\cite{baddeley2015}. Under the assumption of \textit{stationarity}, the properties of the point process are invariant under translation and the process is homogeneous. The \textit{second-order stationarity} states that the second-order intensity only depends on the difference between points $\lambda^{(2)}(\xi_i,\xi_j) = \lambda^{(2)}(\xi_i-\xi_j)$. Because in practice most of processes are inhomogeneous, \cite{baddeley2000,gabriel2009} weakened it and defined the \textit{second-order intensity-reweighted stationary} assumption for which the pair correlation function~(\ref{eq:pcf}) is well-defined and a function of $\xi_i-\xi_j$. \cite{vanlieshout2019}~provides general concepts of factorial moment properties. The previous definition of inhomogeneous processes is not unique, \cite{hahn2015} defined inhomogeneous model classes (including the class of reweighted second-order stationary processes) into the common general framework of hidden second-order stationary processes.
The pair correlation function describes the structure of dependence/interaction between points~:
$g(\xi_i,\xi_j) = 1$, $>1$ and $<1$ indicates that the pattern is, respectively, completely random, clustered and regular.
 
Assume that the distribution of the point process is defined by a \textit{probability density} $f(\textbf{x})$ with respect to the distribution of a unit rate Poisson process. The probability density can be used to study point processes. It can be viewed as the probability of getting the point pattern \textbf{x}, divided by the same probability under Complete Randomness~\cite{baddeley2015}. 
The mathematical form of the probability density determines the structure of the point process, see~\cite{coeurjolly2017,coeurjolly2019} about formulation of the density of point processes.
A closely related concept is the Papangelou conditional intensity function~\cite{papangelou1974}, which has been extended to the spatio-temporal framework by~\cite{cronie2015}. It is defined by
\begin{equation}
\label{eq:papangelou}
\lambda(\xi|\textbf{x}) = \frac {f(\textbf{x} \bigcup  \xi)}{f(\textbf{x})},
\end{equation}
for $\xi \notin \textbf{x}$ provided $f(\textbf{x}) \neq 0$.

\subsection{Classical point process models}

We refer to~\cite{diggle2003,moller2004,illian2008,chiu2013,baddeley2015} 
and~\cite{cronie2015,diggle2010,diggle2013b,gabriel2013,gonzalez2016} for a presentation of most of spatial  and spatio-temporal point process models. Hereafter we only focus on the ones mentioned/used in Section~\ref{sec:methods} to construct multi-structure point process models, namely the Poisson, Cox and Gibbs processes.

\subsubsection*{Poisson point processes}

The Poisson point process is the reference model for independence of the locations of events, i.e. for complete spatial (or spatio-temporal) randomness. It is also the simplest and most widely used inhomogeneous point process model. Poisson point processes with intensity function $\lambda(\xi)$ are defined by two postulates~:
\begin{itemize}
  \item The number of points in any region $B \subseteq {\cal W}$, $N(B)$, follows a Poisson distribution with parameter $\int_{B} \lambda(\xi) \text{d} \xi$,
  \item For all $B \subseteq {\cal W}$, given $N(B) = n$, the $n$ events in $B$ form an independent random sample from the distribution on $B$ with probability density function $\lambda(\xi) / \int_{B} \lambda(\xi) \text{d}  \xi$.
\end{itemize}
The probability density of a Poisson point process with respect to the unit rate Poisson process is
$$f(\textbf{x}) = \exp \left( |{\cal W}| - \int_{\cal W} \lambda(\xi) \text{d} \xi \right) \Pi_{\xi \in \textbf{x}} \lambda(\xi).$$
Then, from Equation~(\ref{eq:papangelou}), the Papangelou conditional intensity is $\lambda(\xi | \textbf{x}) = \lambda(\xi)$ and $\lambda^{(2)}(\xi_i,\xi_j) = \lambda(\xi_i)\lambda(\xi_j)$, so that $g(\xi_i,\xi_j)=1$.

\subsubsection*{Cox processes}

Cox processes, so-called doubly stochastic point processes~\cite{cox1955}, are considered as a generalization of inhomogeneous Poisson processes where the intensity is a realization of a random field $\Lambda= \lbrace \Lambda(\xi) \rbrace_{\xi \in {\cal W}}$.
These models are particularly useful as soon as spatial variation in events density reflects both the environment and dependence between events.
Moreover, their first- and second-order moments being tractable, they are very attractive. We have
\begin{equation}
\label{eq:momentsCox}
  \lambda(\xi) = \mathds{E} \lbrack \Lambda(\xi) \rbrack
  \ \ \text{and} \ \ 
  g(\xi_i,\xi_j) = \frac{\mathds{E} \lbrack \Lambda(\xi_i) \Lambda(\xi_j) \rbrack}{\lambda(\xi_i) \lambda(\xi_j) }
  = 1 + \frac{\text{cov} \left(  \Lambda(\xi_i), \Lambda(\xi_j) \right)}{\lambda(\xi_i) \lambda(\xi_j) }.
\end{equation}
The probability density $f(\textbf{x}) = \mathds{E} \left\lbrack \exp \left( |{\cal W}| - \int_{\cal W} \Lambda(\xi) \text{d}\xi \right) \Pi_{\xi \in \textbf{x}} \Lambda(\xi) \right\rbrack$ is intractable for these processes. Consequently, the Papangelou conditional intensity is not known.
The second-order intensity function $\lambda^{(2)}(\xi_i,\xi_j) =  \mathds{E} \left\lbrack \Lambda(\xi_i) \Lambda(\xi_j) \right\rbrack$
is only tractable for two special cases of Cox processes, that we present below, the \textit{Shot Noise Cox process} and the \textit{log-Gaussian Cox process}.

Shot noise Cox processes~\cite{moller2003} (SNCP) are a wide class of Cox processes associated to 
$$\Lambda(\xi) =\sum_{(c,\gamma) \in \Phi} \gamma k(c,\xi), $$
where $\Phi$ is a Poisson point process on ${\cal W} \times [0,\infty)$ with intensity measure $\zeta$ and $k(c,\cdot)$ is a density function on ${\cal W}$, $\forall c \in {\cal W}$. The intensity and pair correlation function are 
\begin{equation*}
  \lambda(\xi) = \int \gamma k(c,\xi) \text{d}\zeta(c,\gamma) %, \ \forall \xi \in {\cal W}
   \ \ \text{and} \ \
  g(\xi_i,\xi_j) = 1 + \dfrac{\int \gamma^2 k(c,\xi_i) k(c,\xi_j) \text{d}\zeta(c,\gamma)}{\lambda(\xi_i)\lambda(\xi_j)}. % , \ \forall \xi_i \xi_j \in {\cal W}.
\end{equation*}
SNCP include Poisson cluster processes, i.e. a Poisson process in which each point is replaced by a cluster of points, the original point is considered as the cluster center~\cite{cox1980}. When the points in the cluster are independently and identically distributed about the cluster centre, the process is referred to as a Neyman-Scott process~\cite{neyman1958}. Two mathematically tractable models of Neyman-Scott processes are the \textit{Thomas process}~\cite{thomas1949}, where $k$ is a zero-mean normal density, and the \textit{Mat{\'e}rn cluster process}, where $k$ is a uniform density on a ball centered at the origin.

Log-Gaussian Cox processes (LGCP) have been introduced in~\cite{moller1998}, considering that the intensity is a log-Gaussian process~: $\Lambda(\xi) = \exp \left( Y(\xi) \right)$, where $Y$ is a real-valued Gaussian random field, with mean function $\mu(\xi)$ and covariance function $C(\xi_i,\xi_j)$. In that case, from Equation~(\ref{eq:momentsCox}) we have
\begin{equation*}
  \lambda(\xi) = \exp \left( \mu(\xi) + C(\xi,\xi)/2 \right), \ \forall \xi \in {\cal W}
   \ \ \text{and} \ \
  g(\xi_i,\xi_j) = \exp \left( C(\xi_i,\xi_j) \right), \ \forall \xi_i, \xi_j \in {\cal W}.
\end{equation*}
The expression of the pair correlation function shows that the interaction is controlled by the second-order moment of $Y$. If $C(\xi_i,\xi_j) \geq 0$,
we get $g(\xi_i,\xi_j) > 1$ and clustering.
As they are based on a latent random field describing the intensity, LGCPs have a hierarchical structure making them particularly flexible~\cite{illian2008}. Note that the interaction is controlled through the second-order moment of the Gaussian random field, so that LGCPs do not describe the mechanistic process generating the points what is the case of most of Gibbs processes (see below) for which the dependence between points is controlled through local interaction between pairs of points.

\subsubsection*{Gibbs point processes}

A finite Gibbs point process on ${\cal W}$ admits a density 
\begin{equation}\label{eq:gibbs}
f(\textbf{x}) = \exp \left( -\Psi (\textbf{x}) \right)
\end{equation}
 w.r.t. the Poisson process of unit intensity on ${\cal W}$. The potential function $\Psi$ is often specified as the sum of pair potentials~:
\begin{equation}\label{eq:potential}
  \Psi(\xi_1,\dots,\xi_n) = \alpha_0 + \sum_i \alpha_1(\xi_i) + \sum_{i<j} \alpha_2(\xi_i,\xi_j) + \dots + \alpha_n(\xi_1,\dots,\xi_n),
\end{equation}
with $\alpha_0$ a normalizing constant for the density and the pair potentials $\alpha_1, \alpha_2, \dots$ which determine the contribution to the potential from each $\delta$-uple of points. Note that, if the $\alpha_\delta$, $\delta \geq 2$ are identically zero, the process is Poisson with intensity $\lambda(\xi) = \exp( - \alpha_1(\xi))$. Hence, $\alpha_1$ can be viewed as controlling a spatial (or spatio-temporal) trend, while the $\alpha_\delta$, $\delta \geq 2$ control the interactions between events.
The normalizing constant is generally intractable, so it is often impossible to compute the intensity and pair correlation function of Gibbs processes. However, the Papangelou conditional intensity can be computed~\cite{coeurjolly2019}.

When the interaction between points is restricted to pairs, i.e. for
$$f(\textbf{x}) = \alpha \Pi_i \beta(\xi_i) \Pi_{i<j} \gamma(\xi_i,\xi_j),$$
with $\alpha > 0$, $\beta$ an intensity  function and $\gamma$ a symmetric interaction function, the process is called \textit{pairwise interaction process}~\cite{diggle1983,vanlieshout2000}. A well-known example of such processes is the \textit{Strauss process}~\cite{strauss1975} for which
$$f(\textbf{x}) = \alpha \beta^{n(\textbf{x})} \gamma^{s(\textbf{x})}, $$
where $\beta,\gamma>0$, $n(\textbf{x})$ is the number of points in $\textbf{x}$ and $s(\textbf{x})$ the number of neighbour pairs of \textbf{x} at distances less than a given distance $R$. When $\gamma = 0$, we get the \textit{Hard Core process}. Note that in the Strauss process, $\gamma$ should be smaller than 1 otherwise the density is no integrable. 
\cite{geyer1999} modified the Strauss process and proposed the \textit{Geyer saturation process} in which the overall contribution from each point is trimmed to never exceed a maximum value. We thus have
$$f(\textbf{x}) = \alpha \beta^{n(\textbf{x})} \Pi_{\xi \in \textbf{x}} \gamma^{\min(s,t(\xi,r,\textbf{x}))}, $$
where $\alpha,\beta,\gamma,r,s$ are parameters and $t(\xi,r,\textbf{x})$ is the number of other events lying with a distance $r$ of the point $\xi$.

\section{Multi-structure  point process models}
\label{sec:methods}

Spatial and spatio-temporal single-structure  point process models presented in the previous section are generally
used when only one type of interaction governs the structure of the
point pattern. When there are indications that the spatial or spatio-temporal
structure combines several structures or varies with ranges of distances, we need to consider multi-structure point process models. 
% A moi %
We present in this section some of these models derived from the classes of Gibbs and Cox processes.
By nature, few spatial point processes can exhibit directly several structures and/or scales of interaction and we recall some useful construction techniques to incorporate the multi-structure: hybridization, thinning, superposition or clustering.

\subsection{Models based on Gibbs processes}

Gibbs point processes are mainly used to model repulsion structure in point patterns, even if some examples exist for modelling low clustering~\cite{chiu2013}. Their definition through the potential function $\Psi$ fit well in the statistical mechanics framework where the spatial modelling of particles needs often to consider their interaction. It is common in various domains (mechanics, biology$\dots$) to observe repulsion at short range and aggregation at medium-long range of entities, leading to define multi-structure point processes models.

For pairwise interaction processes, some parametric potential functions can be defined to take into account multiple scales of interaction, see e.g. \cite{ruelle1969,ogata1981,penttinen1984,clyde1991,habel2019}. We consider in the sequel the homogeneous case, i.e. when $\alpha_1$ is constant and the pair potential function $\alpha_2(\xi_i,\xi_j)=\alpha_2(\|\xi_i - \xi_j \|)$ in~\eqref{eq:potential}.

The Lennard-Jones pair potential function, well-known in statistical mechanics, is given by
\begin{equation*}
\alpha_2(r) = \epsilon_1 \left(\frac{\sigma}{r}\right)^{m_1} - \epsilon_2 \left(\frac{\sigma}{r}\right)^{m_2}, \quad \forall r > 0
\end{equation*}
where $m_1 > m_2$, $\epsilon_1, \sigma > 0$ and in the multi-structure case $\epsilon_2 > 0$.
Another one is the step potential function given by
\begin{equation*}
\alpha_2(r) = c_l \quad \text{if } R_{l-1} < r \leq R_l \quad \forall l=1,\cdots,m  
\end{equation*}
where $R_0=0$, $R_m=\infty$, $c_1=\infty$, $c_m=0$ and $c_l \in \mathds{R}$ for $l=2,\cdots,m-1$. The resulting model is an extension of the Strauss process to the multi-scale framework \cite{penttinen1984}. The square-well potential is obtained with $l=2$.
More recently, \cite{goldstein2015} introduced a pair potential function varying smoothly over distance with scale interactions defined through a differential system of equations. Other pair potential functions can be found in the literature for modeling multi-structure phenomena, e.g. in \cite{ogata1981,chiu2013}.\\

Some of these pair potential functions define multi-scale generalizations of single scale Gibbs processes.
Indeed, the step potential functions of homogeneous pairwise interaction processes in \cite{diggle1983} and \cite{penttinen1984} represent multi-scale extensions of the Strauss process where the density is given by
\begin{equation*}
f(\textbf{x}) = \alpha \beta^{n(\textbf{x})} \prod_{l=1}^m \gamma_l^{s_l(\textbf{x})}, 
\end{equation*}
where $s_l(\textbf{x}) = \sum_{i<j} \mathbb{1} (R_{l-1} < \| \xi_i - \xi_j \| \leq R_l)$.\\

In the same way, the multi-scale generalization of the area-interaction model has been introduced in~ \cite{ambler2002,ambler2004,ambler2010} with a two-scale structure and in \cite{picard2009} for multi-scale marked area-interaction processes. Its density function in a homogeneous multi-scale case is given by
\begin{equation*}
f(\textbf{x}) = \alpha \beta^{n(\textbf{x})} \prod_{l=1}^m \exp(- \kappa_l U(\textbf{x},r_l))
\end{equation*}
where $U(\textbf{x},r_l)$ is the $d$-dimensional volume of the set ${\cal W} \cap \bigcup_{\xi \in \textbf{x}} b(\xi,r_l) $, with $b(\xi,r_l)$ the ball centered at $\xi_i$ of radius $r_l > 0$. The sign of $\kappa_l$ defines the $l$th structure~: inhibition if negative, clustering otherwise. \cite{nightingale2019} used area-interaction point processes for bivariate point patterns for modelling both attractive and inhibitive intra- and inter-specific interactions of two plant species.\\

\cite{baddeley2013} defined a new class of multi-scale Gibbs point processes named hybrid models and including the two previous generalization examples. This unified framework allows to define properly generalizations of single-scale Gibbs point processes by preserving Ruelle and local stability~\cite{vanlieshout2000}. This hybridization technique consists in defining the density function of a multi-scale point process model as the product of several densities of Gibbs point processes, so that
\begin{equation*}
f(\boldsymbol{x})=cf_1(\boldsymbol{x})...f_m(\boldsymbol{x})
\end{equation*}
where $c$ is a normalization constant and $f_l$ is a Gibbs density function for $l=1,\cdots,m$.
The choice of the normalization constant allows to well define a probability density in the case where the product $f_1...f_m$ is integrable. 
The integrability condition is of course essential and induced by others conditions on the $f_l$ (Ruelle statbility, local stability or hereditary, see~\cite{baddeley2013}) which play an important role in simulation algorithms and are established in general to demonstrate the model validity of the hybrid process.

 %Note that the product of densities of Gibbs point processes does not necessarily define a point process density. Additional assumptions are required to ensure that the hybrid is integrable (see~\cite{baddeley2013}). 

\cite{baddeley2013} introduced the spatial multi-scale Geyer saturation point process that was applied in epidemiology by \cite{iftimi2017}  and in seismology by \cite{siino2017} and \cite{siino2018}. \cite{raeisi2019} extend the definition and the estimation procedure in the general case of an inhomogeneous spatio-temporal multi-scale Geyer saturation process which density is given by
\begin{equation}
f(\textbf{x})=c \prod_{\xi \in \textbf{x}} \lambda(\xi) \prod_{l=1}^{m}\gamma_l ^ {min\{s_l,n(C_{r_l}^{q_l}(\xi);\textbf{x})\}}
\label{eq:densGeyer}
\end{equation}
where $\lambda \geq 0$ is a measurable and bounded function, $\gamma_l, r_l, q_l$ and $s_l >0$ are the model parameters and $n(C_{r}^{q}(\xi);\textbf{x})=\sum _ {\xi_i \in \textbf{x} \setminus \xi}  \mathbb{1} \{||x_i-x||\leq r_l, |t_i-t|\leq q_l\}$ is the number of other points in $\textbf{x}$ which are in a cylinder centred on $\xi \in \textbf{x}$ with spatial and temporal radii $r_l$ and $q_l$. For fixed $l \in \lbrace 1, \dots,m \rbrace$, when $0 < \gamma_l < 1$ we would expect to see inhibition between events at spatio-temporal scales. On the other hand, when $\gamma_l > 1$ we expect clustering between events. We observe that Equation~\eqref{eq:densGeyer} reduces to an inhomogeneous Poisson process when  $s_l = 0$ $\forall l \in \lbrace 1, \dots,m \rbrace$. \cite{rajala2018} used a multitype generalization of Gibbs point processes with point-to-point interactions at different spatial scales in order to model a complex rainforest data of $83$ species. 

The definition of hybrid Gibbs models does not impose to consider the same $m$ Gibbs models which is emphasized in \cite{baddeley2015}. In this way, \cite{badreldin2015} applied a hybrid model with three model structures at different ranges of distance to the spatial pattern of halophytic species distribution in an arid coastal environment. They considered a hardcore process at very short distances, a Geyer process at short to medium distances and a Strauss process for the structure at large distances.

\subsection{Models based on Cox processes}

Cox processes are mainly defined from additive or log-linear random intensity functions. Their hierarchical structure allows to quantify the various sources of variation governing the spatial or spatio-temporal distribution of the pattern of interest. They are widely used for modelling environmental and ecological patterns.

\subsubsection*{Cluster Cox processes and superposition}

Some Cox processes are obtained by clustering of \textit{offspring} points around \textit{parent} points and correspond to specific cases of cluster processes. This two-step construction allows to consider easily different structures for the patterns of parents and offspring.

\cite{moller2005} introduced the class of Generalized Shot Noise Cox processes (GSNCP), extending the definition of SNCP, and allowing relevant multi-structure point processes for modelling regularity and clustering in many applications. This class has two advantages. Firstly, the parent process is not restricted to be Poisson, as in Neyman-Scott processes, and can be a repulsive Gibbs point process in order to add inhibition between the clusters. Secondly, in each cluster, the intensity and the bandwidth of the dispersion kernel can be random. By consequence, a GSNCP is a Cox process driven by a random field of the form
\begin{equation*}
\Lambda(\xi) = \sum_{(c,\gamma,h) \in \Phi} \gamma k_h(c,\xi),
\end{equation*}
where $\Phi$ is a point process on $\mathcal{W} \times [0,\infty) \times [0,\infty)$ and $h$ is a bandwidth for the kernel density $k_h(c,\cdot)$.
So, given $\Phi$, a GSNCP is distributed as the superposition $\cup_{l} X_l$ of independent Poisson processes with intensity functions $\gamma_l k_{h_l}(c_l,\cdot)$ where $\{\gamma_l\}_l$, $\{h_l\}_l$ are random and $\Phi_{cent} = \{c_l\}_l$ is the parent process.
In population dynamics, with $G_0$ a Poisson process for the initial population and $G_{n+1}$ a GSNCP where the cluster centers are given by $G_n$, the superposition of GSNCPs $G_0, G_1, \dots$ is a spatial Hawkes process~\cite{hawkes1971}.
The GSNCP class contains the special cluster Cox process defined in \cite{Yau2012}, where the parents process is a Strauss process. This model coupling inhibition at medium/long range and aggregation in cluster is applied to tree locations in a rain-forest, in order to consider the competition and reproduction mechanisms. 
\cite{Albert-Green2016} and \cite{Albert-Green2019} generalized the Neymann-Scott process by considering a log-Gaussian Cox process model for the parents, instead of a homogeneous Poisson process, leading to two scales of clustering, inter- and intra-clusters. This hierarchical model is applied to storm cell modelling in North Dakota.

Wiegand and co-authors' papers~\cite{wiegand2007,wiegand2009} consider several construction of Cox processes incorporating clustering at multiple scales. The nested double-cluster process is an extension of the Thomas process in an multi-generation evolution of the population where the offspring become parents and generate offspring. They consider also the superposition of cluster processes, like the Thomas process.

\subsubsection*{Cox processes with constructed covariate}

Another way to incorporate both small and large spatial scale structure in Cox processes is to define a constructed covariate measuring the local structure of a point pattern associated to an additional spatial effect at medium-long range. This methodology developed in~\cite{illian2012a} and applied to koala data is used again in~\cite{illian2012b,illian2013} for other spatial ecological data. They consider a log-Gaussian Cox process in a Bayesian framework in order to apply the INLA approach for speeding up the estimation of parameters in comparison to MCMC approaches that are very time-consuming. \cite{gabriel2017} used also this approach in the context of wildfire modelling in Mediterranean France. In the case of a spatial LGCP model, the method consists in estimating the random field $\Lambda$ on grid cells $s_i$ as follow 
\begin{equation*}
\Lambda(s_i) = \exp \left( \beta_0 + f(z_c(s_i)) + \sum_{k=1}^p f_{k}(z_k(s_i)) + Y(s_i) \right)
\end{equation*}
where $\beta_0$ is the intercept, $f(z_c(\cdot))$ is a function of the constructed covariate $z_c$, $f_k, k=1,\cdots,p$ are functions of the observed covariates $z_k$ and $Y$ is a Gaussian random field taking into account the spatial autocorrelation not explained by the covariates. This intensity is estimated for each cell $s_i$ of a grid partitioning the observation window.

In \cite{illian2012a}, the constructed covariate at each center point $c$ of the grid cell $s$ is the distance from $c$ to the nearest point in the pattern outside the grid cell, i.e $z_c(s) = \min_{\xi \in \textbf{x} \setminus s} (\|c - \xi\|)$.
This constructed covariate describes small scale inter-individual behavior whereas the random field $Y$ captures the spatial autocorrelation at a large spatial scale. The space-time and space-mark extensions of the constructed covariate definition are respectively introduced in \cite{illian2012b} and \cite{illian2013}. In \cite{gabriel2017} the constructed covariate corresponds to a temporal intensity index given by the ratio between the number of wildfires observed spatially close to an other in a specified period and the total number of closed wildfires observed outside this given period. This covariate measures the temporal wildfire inhibition at close spatial distances induced by the local burn of vegetation after a wildfire occurrence.
\cite{sorbye2019} fitted a LGCP to rainforest tree species by adding to the combination of covariates in the log-intensity a spatial random field and error field. The first random field captures the spatial autocorrelation in point counts among neighboring grid cells and the second one the clustering within grid cells, as a nugget effect in geostatistics. The intensity in $s \in \mathcal{W}$ is thus given by
\begin{equation*}
\Lambda(s) = \exp \left( \beta_0 + \sum_{k=1}^p \beta_k z_k(s) + \frac{1}{\sqrt{\tau}} \left\{ \sqrt{\rho} \times Y(s) + \sqrt{1-\rho} \times \epsilon(s) \right\} \right)    
\end{equation*}
where $\beta_k$ are linear effects of observed covariates $z_k$, $Y$ is a spatial random field with autocorrelation between grid cells and $\epsilon$ the error field driving the aggregation structure within grid cells.

\subsubsection*{Thinned point processes}

Thinning is a an operation allowing to delete points in a point process in order to obtain a new one with different characteristics. Each point of a point process has a probability $1-\pi$ of deletion, where the retention probability $\pi$ can be constant or not, independent of the location point or depending on one to several points. For Cox processes, this technique is generally applied to create random local regularity. For example, \cite{andersen2016} applied a Mat{\'e}rn hard core dependent thinning to a Shot Noise Cox process to obtain short range repulsion with medium range clustering. For a given point pattern and
a specified distance $h$, Mat{\'e}rn hard core thinning acts by first attaching random positive marks (arrival times) to each point. Subsequently a point is removed if it
has a neighbour within distance $h$ and with a smaller mark (i.e. the neighbour
arrived earlier). In that way, for a given location $\xi$, the retention probability $\pi(\xi)$ is the ratio between the intensities of the thinned
process and the original process at $\xi$.
\cite{lavancier2016} extended the definition of interrupted point processes in \cite{stoyan1979} and \cite{chiu2013} and considered a spatial point process $X$ obtained by an independent thinning driven by a random process $Z$ on a regular point process $Y$. An example is given with $Y$ a Mat{\'e}rn hard core process and $Z$ the transformation by a characteristic function of a Boolean disc model~\cite{chiu2013}.

\section{Discussion and conclusion}
\label{sec:discussion}

This paper presents a review of methods for constructing multi-structure point processes for modelling  aggregation and/or inhibition at different spatial or spatio-temporal scales. We focus our attention on the main two classes of point processes, namely the Gibbs and Cox processes. Some multi-structure techniques are specific to a family of point processes, as the hybridization approach for Gibbs processes or the double-cluster process for Cox processes; others are more global, as the superposition or the thinning method, even if they are respectively more adapted to Gibbs or Cox processes. We could also consider determinantal point processes to model regularity as in \cite{lavancier2016} who considered it instead of the Mat{\'e}rn hard core process. % for the initial unobserved regular process $Y$. 
Spatio-temporal point processes can also be defined by conditioning on the past, often used in epidemiology or seismology. For instance, the definition of the conditional intensity in~\cite{diggle2006} allows an aggregation of cases in the spatio-temporal spread of the foot and mouth disease and also a random occurrence of cases in the entire observation domain.

We selected the most relevant references for us in the state-of-the-art of these types of Gibbs and Cox models to describe these approaches for introducing regularity in cluster processes and aggregation in repulsive processes. Because these models are suitable in an environmental and ecological framework, due to the complexity of mechanisms governing attraction and repulsion of entities (particles, cells, plants$\dots$), we can expect a wide use of these models in many studies.\\

\end{document}